\documentclass[12pt]{article}

\usepackage{amsmath,amsfonts,amssymb,latexsym}

\newtheorem{theorem}{Theorem}

\numberwithin{equation}{section}
\def\be{\begin{equation}}
\def\ee{\end{equation}}
\def\bq{\begin{eqnarray}}
\def\eq{\end{eqnarray}}
\def\beq{\begin{eqnarray*}}
\def\eeq{\end{eqnarray*}}

\begin{document}
\begin{titlepage}
\begin{flushright}
\end{flushright}

\vspace{0.7cm}

\begin{center}
{\huge Modern Approaches to Cosmological Singularities}

\vspace{1cm}

{\large Spiros Cotsakis$\dagger$ and Ifigeneia Klaoudatou$\ddagger$}\\

\vspace{0.5cm}

{\normalsize {\em Research Group of Cosmology, Geometry and
Relativity}}\\ {\normalsize {\em Department of Information and
Communication Systems Engineering}}\\ {\normalsize {\em University
of the Aegean}}\\ {\normalsize {\em Karlovassi 83 200, Samos,
Greece}}\\ {\normalsize {\em E-mail:}
$\dagger$\texttt{skot@aegean.gr},
$\ddagger$\texttt{iklaoud@aegean.gr}}
\end{center}

\vspace{0.7cm}

\begin{abstract}
\noindent We review recent work and present new examples about the
character of singularities in globally and regularly hyperbolic,
isotropic universes. These include  recent singular relativistic
models, tachyonic and phantom universes as well as  inflationary
cosmologies.
\end{abstract}

\vspace{0.3cm}
\begin{center}
{\line(5,0){280}}
\end{center}

\end{titlepage}

\section{Introduction}
An important question eventually arising in every study of the
global geometric and physical properties of the universe is that
of deciding whether or not the resulting model is geodesically
complete. Geodesic completeness is associated with an infinite
proper time interval of existence of privileged observers and
implies that such a universe will exist forever. Its negation,
geodesic incompleteness or the existence of future and/or past
singularities of a spacetime, is often connected to an `end of
time' for the whole universe modeled by the spacetime in question.
By now there exist simple singular cosmological models of all
sorts, not incompatible with recent observations, in which an
all-encompassing singularity features as such a catastrophic
event.

It is well known that in general relativity there are a number of
rigorous theorems predicting the existence of spacetime
singularities in the form of geodesic incompleteness under certain
geometric and topological conditions (see, e.g.,  \cite{cot02} for
a recent review). These conditions can be interpreted as
restrictions on the physical matter content  as well as
plausibility assumptions on the causal structure  of the spacetime
in question. Because all these assumptions are not unreasonable,
the singularity theorems predicting the existence of spacetime
singularities in cosmology and gravitational collapse have become
standard ingredients of the current cosmological theory.

However, such existence results cannot offer any clue about the
generic nature of the singularities they predict.  In addition,
there are new completeness theorems (cf. \cite{cho-cot}) which say
that under equally general geometric assumptions, generic
spacetimes are future (or past) geodesically complete. Among the
chief hypotheses of the completeness theorems, except the usual
causality ones also present in the singularity theorems,  is the
assumption that the space slice does not `vibrate' too much as it
moves forward (or backward) in time, and also the assumption that
space does not curve itself too much in spacetime. It may well be
that cosmological models in a generic sense are only mildly
singular or even complete. It remains thus a basic open problem to
decide about the character of the cosmological singularities
predicted by the singularity theorems.

In preparing to tackle such basic open questions more information
is required, and one feels that perhaps different techniques are
needed which the singularity theorems cannot provide. We are
therefore faced with the following problem: Suppose we have a
spacetime which \emph{is known} to have a singularity. How can we
unravel its basic characteristics and find criteria classifying
different singular spacetimes? What methods are to be used in such
pursuits? Indeed, how are we to make a start into the `zoology' of
cosmological singularities? Such questions apply with equal
interest to different classes of cosmological models (classified
according to symmetry), as well as to the general case.

In this paper, we review recent work on this subject contained
mainly in Refs. \cite{cho-cot}-\cite{cot-kla04} about the
character of cosmological singularities. We present a basic
theorem providing necessary conditions for singularities and give
new examples to illustrate this result. These examples are
constructed using relativistic cosmological models, phantom
cosmologies and inflationary models.

\section{Completeness and the character of singularities}
Although we focus below exclusively on isotropic models,  it is
instructive to begin our analysis by taking a more general stance.
Consider a \emph{slice space} (cf. \cite{co03} for this
terminology), that is a spacetime $(\mathcal{V},g)$ with
$\mathcal{V}=\mathcal{M}\times \mathcal{I},\;$ $\mathcal{I}
=(t_{0},\infty )$, where $\mathcal{M}$ is a smooth manifold of
dimension $n$ and $^{(n+1)}g$ a Lorentzian metric which in the
usual $n+1$ splitting, reads
\begin{equation}
^{(n+1)}g\equiv -N^{2}(\theta ^{0})^{2}+g_{ij}\;\theta ^{i}\theta
^{j},\quad \theta ^{0}=dt,\quad \theta ^{i}\equiv dx^{i}+\beta
^{i}dt.  \label{2.1}
\end{equation}
Here $N=N(t,x^{i})$ is the \emph{lapse function}, $\beta
^{i}(t,x^{j})$  the \emph{shift function} and the spatial slices
$\mathcal{M}_{t}\,(=\mathcal{M}\times \{t\})$ are spacelike
submanifolds endowed with the time-dependent spatial metric
$g_{t}\equiv g_{ij}dx^{i}dx^{j}$. We assume that $(\mathcal{V},g)$
is \emph{globally hyperbolic} and so time-oriented by increasing
$t$. \footnote{We choose $\mathcal{I} =(t_{0},\infty )$ because we
study the future singularity behaviour of an expanding universe
with a singularity in the past, for instance at $t=0<t_{0}$.
However, since $t$ is just a coordinate, our study could apply as
well to any interval $\mathcal{I}\subset\mathbb{R}$.} We also
assume that $(\mathcal{V},g)$ is \emph{regularly hyperbolic}
meaning that the lapse, shift and spatial metric are uniformly
bounded. It is known that a regularly hyperbolic spacetime is
globally hyperbolic if and only if each slice is a complete
Riemannian manifold, cf. \cite{co03}.

A Friedmann universe is a sliced space with $N=1$, $\beta =0$ and
the spatial metric is described by a single function of the
(proper) time, the \emph{expansion scale factor} $a(t)$. Thus the
metric has the form $ds^2=-dt^2+a^2(t)d\sigma^2$ with $d\sigma^2$
denoting the time-independent slice metric of constant curvature
$k$, and the \emph{Hubble expansion rate} is proportional to the
extrinsic curvature of the slices, ${|K|_{g}}
^{2}=3({\dot{a}/a})^{2}=3H^{2}$. Using the completeness theorems
of \cite{cho-cot} we arrive at the following result for isotropic
cosmologies.
\begin{theorem}[Completeness of Friedmann
universes]\label{frwcomp} Every globally hyperbolic, regularly
hyperbolic Friedmann solution such that for each finite $t_{1}$
the Hubble expansion rate $H(t)$ is bounded by a function of $t$
which is integrable on $ [t_{1},+\infty )$,  is  future timelike
and null geodesically complete.
\end{theorem}

We can further use this result to arrive at a characterization of
the different singularities that may arise in isotropic universes.
Consider a singular, globally and regularly hyperbolic (scale
factor assumed bounded only below in this case) Friedmann
universe. Then according to Theorem \ref{frwcomp}, there is a
finite time $t_1$ for which $H$ fails to be integrable on the
proper  time interval $[t_1,\infty)$. In turn, this
non-integrability of the expansion rate $H$ can be implemented in
different ways and we arrive at the following result for the types
of future singularities that can occur in isotropic universes (see
\cite{cot-kla04}).
\begin{theorem}[Character  of future singularities]
\label{2} Necessary conditions for the existence of future
singularities in globally hyperbolic, regularly hyperbolic
Friedmann universes are:
\begin{description}
\item[S1] For each finite $t$, H is non-integrable on $[t_1,t]$,
or
\item[S2] H blows up in a finite time, or
\item[S3] H is defined and integrable (that is bounded, finite) for only a finite
proper time interval.
\end{description}
\end{theorem}

The character of the singularities in this theorem is expected to
be in a sense somewhat milder than standard all-encompassing
big-crunch type ones predicted by the Hawking-Penrose singularity
theorems. For instance, those satisfying condition $S1$ may
correspond to `sudden' singularities located at the right end (say
$t_s$) at which  $H$ is defined and finite but \emph{the left
limit}, $\lim_{\tau\rightarrow t_1^+}H(\tau)$, may fail to exist,
thus making $H$ non-integrable on $[t_1,t_s]$, for \emph{any}
finite $t_s$ (which is of course arbitrary but fixed from the
start). We shall see examples of this behaviour in the next
Section. Condition $S2$ leads to what is called here a blow-up
singularity corresponding  to a future singularity characterized
by a blow-up in the Hubble parameter\footnote{Note that $S1$ is
not implied by $S2$ for if $H$ blows up at some finite time $t_s$
after $t_1$, then it may still be integrable on $[t_1,t]$,
$t_1<t<t_s$.}. Condition $S3$  may also lead to a singularity but
for this to be a genuine species (in the sense of geodesic
incompleteness) one needs to demonstrate that  the metric is
non-extendible to a larger interval.

\section{General relativistic isotropic models}
We now consider an example of a cosmological model with a future
singularity which fulfills precisely our condition $S1$ of the
previous Section. This model, as given by Barrow in his definite
recent works \cite{ba04a,ba04b}, is described by the most general
solution of the Friedmann equations for a perfect fluid source
with equation of state $p=w\rho$ in a local neighborhood of the
singularity which is located at the time $t=t_s$ ahead:
\begin{equation}\label{barrow}
a(t)=1+\left(\frac{t}{t_{s}}\right)^{q}
(a_{s}-1)+\tau^n\Psi(\tau),\quad \tau=t_s-t.
\end{equation}
Here we take $1<n<2,\, 0<q< 1, a(t_s)=a_s$ and $\Psi(\tau)$ is the
so-called \emph{logarithmic psi-series} which is assumed to be
convergent, tending to zero as $\tau\rightarrow 0$. Barrow shows
in \cite{ba04a,ba04b} that the form (\ref{barrow}) exists as a
\emph{smooth} solution only on the interval $(0,t_{s})$. Also
$a_{s}$ and $H_{s}\equiv H(t_s)$ are finite at the right end but
$\dot{a}$ blows up as $t\rightarrow 0$ making $H$ continuous only
on $(0,t_s)$. In addition,  $a(0)$ is finite and we can extend $H$
and define it to be finite also at $0$, $H(0)\equiv H_0$, so that
$H$ is defined on $[0,t_s]$. However, since $\lim_{t\rightarrow
0^+}H(t)=\pm\infty$, we conclude that this model universe
implements exactly Condition $S1$ of the previous Section and thus
$H$ is non-integrable on $[0,t_s]$, $t_s$ arbitrary.

This then provides an example of the so-called  big rip
singularity characterized by the fact that as $t\rightarrow t_{s}$
one obtains $\ddot{a}\rightarrow-\infty$. Then using the field
equation we see that this is really a divergence in the pressure,
$p\rightarrow\infty$. In particular, we cannot have in this
universe a family of privileged observers each having an infinite
proper time and finite $H$. A further calculation shows that the
product $E_{\alpha\beta}E^{\alpha\beta}$, $E_{\alpha\beta}$ being
the Einstein tensor, is unbounded at $t_s$. Hence we find that
this spacetime is geodesically incomplete.

\section{Tachyonic cosmologies}
Tachyons, phantoms, Chaplygin gases and quintessence represent
unobserved and unknown, tensile, negative energy and/or pressure
density substances, violating some or all of the usual energy
conditions, whose purpose is to cause cosmic acceleration and
drive the late phases of the evolution of the universe (see
\cite{gib03,sen03} and references therein). Are such universes
generically singular or complete? Such scalar sources usually have
the unpleasant property of super-luminal sound speed at low
density (an exception to this rule is given in \cite{gib01}) and
the counter-intuitive property of the sound speed going to zero at
large density.

Despite these negative features, one can easily construct complete
as well as singular models (see \cite{cot-kla04} for more
examples of this sort). Consider a Friedmann universe filled
with a generalized Chaplygin gas with equation of state given by
\cite{go03}
\begin{equation}
p=-{\rho}^{-\alpha}[C+(\rho^{1+\alpha}-C)^{\alpha/(1+\alpha)}],
\end{equation}
where $C=A/(1+w)-1$ and subject to the condition
$1+\alpha=1/(1+w)$. The scale factor is given by the form
\begin{equation}
a(t)=\left(C_{1}e^{-C_3\tau}
+C_{2}e^{C_3\tau}\right)^{2/3},\quad\tau=t-t_0,
\end{equation}
where $C_{1}$, $C_{2}$ and $C_{3}$ are constants. Therefore we find that in the asymptotic
limits  $\tau\rightarrow 0$ and $\tau\rightarrow\infty$,
$H$ tends to suitable constants, that is it
remains finite on $[t_{0},\infty)$ and  the model is geodesically
complete.

Completeness, however,  is really a property sensitive
to the equation of state assumed in any particular model. If
instead one assumes that the dark energy component satisfies at
late times a general equation of state of the form \cite{noj05}
\begin{equation}
p=-\rho-f(\rho),\quad f(\rho)=A\rho^{\alpha},
\end{equation}
with $A$ and $\alpha$ being real parameters, then setting
$\bar{A}=A\rho_{0}^{\alpha-1}>0$ one finds that for $\alpha\in
(1/2,1)$ the Hubble expansion rate becomes
\begin{equation}
H=C\left({\left(1+3\bar{A}(1-\alpha)
\ln\frac{a_{d}}{a_{0}}\right)}^{\frac{1-2\alpha}{2(1-\alpha)}}+
\frac{3}{2}\bar{A}(1-2\alpha)C(t-t_{d})\right)^{\frac{1}{1-2\alpha}},
\end{equation}
where $t_d$ is the time when dark energy dominance commences.
Therefore at the time $t_f$, where
\begin{equation}
\label{eqH} t_{f}=t_{d}+\frac{2}{3\bar{A}(2\alpha-1)C}
{\left(1+3\bar{A}(1-\alpha)\ln{\frac{a_{d}}{a_{0}}}\right)}^{(1-2\alpha)/(2(1-\alpha))},
\end{equation}
$H$ blows up. This is clearly a type-$S2$ singularity according to
Theorem \ref{2}. In contrast, when $\alpha<1/2$ we
see from (\ref{eqH}) that $H$ is always finite and therefore the
model is geodesically complete.

It is clear that in the field of phantom cosmology more work is
needed to decide on the issue of future singularities.

\section{Inflationary models}
Inflation continuously produces thermalized regions coexisting
with ones still in an inflationary phase and so the universe,
assumed to consist generically of regions of both types,
\emph{must} be future geodesically complete (`future eternal'
according to the inflationistic terminology, cf. \cite{vil,lin}).
It is an intriguing question for every inflationary model whether
or not it is eternal in the past direction. Assume for the moment
that it is not past eternal, so the universe is geodesically
incomplete to the past. Then from Theorem \ref{2} one expects that
$H$ will be non-integrable in some way, e.g., according to one of
the conditions  $S1-S3$. Indeed, in Ref. \cite{bo01}, it was shown
that inflation is past singular in the sense of $S3$. Suppose we
are in a flat Friedmann universe and that $H$ is integrable on a
finite interval $[t_i,t_f]$. We assume that the mean of
$H(\lambda)$ on $[t_i,t_f]$, $H_{av}$, satisfies the
\emph{averaged expansion condition} \cite{bo01}
\be
H_{av}>0, \ee along some null geodesic with affine parameter
$\lambda$. Then one finds that
\begin{equation}\label{7.5}
0<\left(\lambda(t_{f})-\lambda(t_{i})\right)H_{av}<\infty,
\end{equation}
which means  that the affine parameter of this past-directed null
geodesic
 must take values only in a finite interval. This in turn says
 that this geodesic is incomplete.
 A similar proof is obtained for the case of a timelike
geodesic. Observe that condition (\ref{7.5}) holds if and only if
the hypotheses in Theorem \ref{frwcomp} are valid for a finite
interval of time \emph{only}, thus leading to incompleteness
according to Theorem \ref{2}, condition $S3$. A similar bound for
the Hubble parameter is obtained in \cite{bo01}  for the general
case, and one therefore concludes that such a model must be
geodesically incomplete.

As expected from Theorem \ref{frwcomp}, a relaxation of the
requirement that  $H$ be finite for only a finite amount of proper
time leads, to singularity-free inflationary models evading the
previously encountered singularity behaviour. Such models have
been preliminary considered in \cite{cot-kla04}.

Space does not allow us to consider the singularity problem in
many interesting recent quantum cosmological models. We leave such
matters to a future publication.

\section*{Acknowledgements} This work was supported by the joint
Greek Ministry of Education/European Union Research grants
`Pythagoras' No. 1351, `Heracleitus' No. 1337, and by a State
Scholarship Foundation grant, and this support is gratefully
acknowledged. The work of I.K. is for the partial fulfillment of
the PhD thesis requirements, University of the Aegean.

\end{document}